\documentstyle[12pt,psfig]{article}
\def\reference{\parskip 0pt\par\noindent\hangindent 0.5 truecm}

\begin{document}
%
%
\title{Low mass AGB stellar models for $0.003 \le Z \le 0.02$: 
basic formulae for nucleosynthesis calculations}
%
\author{O. Straniero$^{1}$,
 I. Dom\'\i nguez$^{2}$,
 S. Cristallo$^{3}$,
 R. Gallino$^{4}$
} 
\date{}
\maketitle
{\center $^1$ INAF, Osservatorio Astronomico di Teramo, Italy\\ straniero@te.astro.it\\[3mm]
$^2$ Universidad de Granada, Spain\\ inma@ugr.es\\[3mm]
$^3$ INAF, Osservatorio Astronomico di Teramo, Italy\\ cristallo@te.astro.it\\[3mm]
$^4$ Dip. Fisica Generale Universit\`a di Torino and Sez. INFN Torino, Italy\\ gallino@ph.unito.it\\[3mm]
}
\begin{abstract}
We have extended our published set of low mass AGB stellar models to lower metallicity.
Different mass loss rates have been explored. Interpolation formulae for 
luminosity, effective temperature, core mass, mass of dredge up material 
and maximum temperature in the convective zone generated by thermal 
pulses are provided. Finally, we discuss the modifications of these quantities as obtained 
when an appropriate treatment of the inward propagation of the convective instability, 
caused by the steep rise of the radiative opacity 
occurring when the convective envelope penetrates 
the H-depleted region, is taken into account.
\end{abstract}

\bigskip
%
%
\section{Introduction.}

Results from detailed AGB stellar models demonstrate that 
low mass AGB stars (i.e. $1.3 \le M/M_\odot \le 3$) 
are the producers of the main component of the 
cosmic $s$-elements (Straniero et al. 1995; Gallino et al. 1998; Busso et al. 1999).
Such a scenario has been confirmed by the 
measurements of the chemical composition of AGB stars 
(Lambert  et al. 1995; Busso et al. 2001; Abia et al. 2001,2002) and by
the analysis of the isotopic composition of meteoritic SiC grains 
(Gallino et al. 1997). 

The present generation of AGB stars, as
 observed in the disk of our Galaxy, have a nearly solar chemical composition. However, 
SiC grains found in pristine meteorites originated in the C-rich circumstellar envelope 
of a pre-solar generation of AGB stars, whose original metallicity could have been 
somewhat lower than that found in the solar system material. 
In addition, an important contribution to our understanding of AGB stars comes from
observations of the stellar population in the fields of the 
Small and the Large Magellanic Clouds, 
whose average metallicities are about 1/5 and 1/2 of the solar, respectively.
In a previous paper (Straniero et al. 1997) 
the properties of low mass AGB stellar models 
with solar composition have been extensively discussed. Here we present an extension 
of these models to lower metallicity. 

\section{The grid of models and interpolations.}

The full grid of models is summarized in Tab. 1. The corresponding AGB evolutionary
sequences have been obtained by means of the FRANEC code, the
same described in Straniero et al. (1997). The initial mass, the helium content and the metallicity
are reported in column 1, 2 and 3, respectively.  
Various mass loss rates have been applied during the AGB phase. In all 
 cases, we have used the Reimers formula (Reimers, 1975): 
 
\bigskip
\noindent
$\dot M (M_\odot/yr) = 1.34 \cdot 10^{-5} \cdot \eta \cdot \frac {L^{3 \over 2} } {M \cdot T_{eff}^2}$ 

\bigskip
\noindent

where $M$ and $L$ are in solar units and $\eta$ is a free parameter (see fourth column in Tab. 1). 
  We have 
not considered mass loss in pre-AGB evolution, since, in the mass range 1.5-3 M$_\odot$, 
just a few hundredths of solar masses are expected to be lost. 
 For example,  by adopting  $\eta$=0.4 as representative of the pre-AGB mass loss rate 
 (e.g. Fusi Pecci \& Renzini 1976), 
we found that the total mass lost before the onset of the AGB phase by a 2 M$_\odot$ star with solar 
   composition should be 0.04 M$_\odot$. Clearly, such a small amount of mass loss has negligible 
   consequences on the pre-AGB evolution.

The thermally pulsing AGB models (TP-AGB) are characterized by three distinct phases:

1) The {\it early phase}. It includes the first few (gentle) thermal pulses which 
are not followed by the third dredge up (TDU); 

2) The {\it TDU phase}. It begins when the mass of the H exhausted core becomes
as large as $0.59-0.6 M_\odot$. The amount of mass dredged up
firstly increases, as the core mass increases, and then decreases, 
when the envelope mass is substantially reduced;

3) The {\it final phase}. During the last part of the AGB, 
the concurrent actions of mass loss and H-burning erode the envelope mass
and the TDU ceases. The minimum envelope mass for the occurrence of the TDU is
about $0.4-0.5 M_\odot$.

The variation of the mass dredged up, 
as a function of the core mass, is shown in Fig. 1 for some selected sequences of models.

\begin{figure}
\begin{center}
{\centerline{\psfig{file=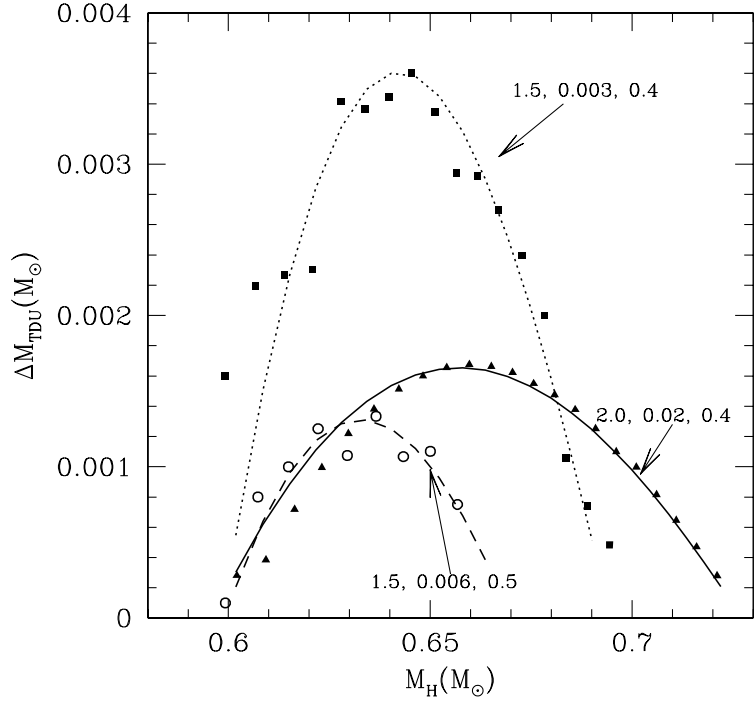,width=12cm}}}
\caption{The amount of mass dredged up per thermal pulse for three evolutionary sequences: $M=2.0$ $M_\odot$, $Z=0.02$ and 
$\eta=0.4$ - triangles (FRANEC models) and solid line (interpolation formula); $M=1.5$ $M_\odot$, $Z=0.006$ and 
$\eta=0.5$ - circles (FRANEC models) and dashed line (interpolation formula); $M=1.5$ $M_\odot$, $Z=0.003$ and 
$\eta=0.4$ - squares (FRANEC models) and dotted line (interpolation formula).}
\end{center}
\end{figure}

An interpolation on the grid of models provides useful analytic expressions for
relevant quantities, as a function of $M_{\rm H}$ (the mass of the H-exhausted core, in 
$M_\odot$), $M_{env}$ (the mass of the H-rich envelope, in $M_\odot$) and Z (the initial mass
fraction of elements with $A \ge 12$). 
         
In the following we report the result of these interpolations.

\bigskip
\noindent
1. The time elapsed between two successive thermal pulses (interpulse period): 

\bigskip
$\Delta t_{ip}(10^5 {\rm yr})=18.5178-59.8876 \cdot M_{\rm H}+48.8462 \cdot M_{\rm H}^2-4.0273 \cdot \log {Z}+$

\bigskip
\noindent
$+5.8422 \cdot M_{\rm H} \cdot \log {Z}$

\bigskip
\noindent
which is valid for $M_{\rm H} \ge 0.58$. For smaller core masses, 
substitute $M_{\rm H}$ with $(1.16-M_{\rm H})$. The standard deviation of this fit (i.e. the average 
 difference between the values calculated with the fitting formula and those of the models) is 0.01.

\bigskip
\noindent
2. The mass of the H-depleted material dredged up in a given TDU episode:

\bigskip
$\Delta M_{TDU}(M_\odot)=(1.0 + 0.21 \cdot \log {Z \over Z_\odot} + 6.3 \cdot (\log {Z \over Z_\odot})^2)(-6.2 \cdot 10^{-4}+$

\bigskip
\noindent
$+9.1 \cdot 10^{-4} \cdot M_{env} -3.7 \cdot 10^{-4} \cdot M_{env}^2+5.061 \cdot 10^{-2} \cdot M_{env} \cdot \delta M_{\rm H}-$

\bigskip
\noindent
$-5.96 \cdot 10^{-3}
\cdot M_{env}^2 \cdot \delta M_{\rm H}-3.8372 \cdot 10^{-1} \cdot M_{env} \cdot \delta M_{\rm H}^2+9.448 \cdot 10^{-2} \cdot M_{env}^2 \cdot \delta M_{\rm H}^2)$

\bigskip
\noindent
where $Z_\odot=0.02$ and $\delta M_{\rm H}=M_{\rm H}-M_{\rm H}^*$ and $M_{\rm H}^*$ is the core mass at the last thermal 
pulse without TDU (last pulse of the {\it early phase}). 
 For the masses and chemical compositions we have considered, 
$M_{\rm H}^*$ ranges between 0.59 and 0.6. The curves reported in Fig. 1 have been obtained by
using  $M_{\rm H}^*=0.595$.  
When $M_{env}<0.4$ or $\delta M_{\rm H} \le 0$, take $\Delta M_{TDU}=0$. The standard deviation of this 
fitting formula is 2 $\cdot$ 10$^{-4}$.

\bigskip
\noindent
3) The luminosity at half of the interpulse period:

\bigskip
$\log {L \over L_\odot}=3.603+6.7806 \cdot \Delta M_{\rm H} -27.582 \cdot \Delta M_{\rm H}^2+277.333 \cdot \Delta M_{\rm H}^4$

\bigskip
\noindent
here $\Delta M_{\rm H}=M_{\rm H}-M_{\rm H}(t_0)$ and $M_{\rm H}(t_0)$ is the core mass 
at the epoch of the first thermal pulse ($t_0$). The latter quantity slightly depends on the metallicity:
$M_{\rm H}(t_0)=0.517-1.939 10^{-2} \times \log {Z}$. 
 For this fit the standard deviation is 10$^{-4}$.

\bigskip
\noindent
4) The effective temperature at half of the interpulse period:

\bigskip
$\log {T_{eff}} (K)=5.0475 + 0.1438 \cdot \log Z_{eff} + 0.0513 \cdot (\log Z_{eff})^2 - 4.1895 \cdot M_{\rm H} + 2.9594 \cdot M_{\rm H}^2$ 

\bigskip
\noindent
here, $Z_{eff}$ is the effective metallicity, which includes the extra-carbon
in the envelope due to the third dredge up.
The use of $Z_{eff}$, instead of $Z$, has minor effects on solar composition stars, but
may significantly increase the estimated radius and, in turn, the mass loss rate
for stars with an original low content of metal. The standard deviation is 5 $\cdot$ 10$^{-4}$.  

\bigskip
\noindent
5) The maximum temperature attained at the bottom of the convective zone
 generated by a thermal pulse:

\bigskip
$\log {T_{max}} (K) = 6.7747 + 4.6856 \cdot M_{\rm H} - 3.21 \cdot M_{\rm H}^2 + 0.01 \cdot M_{env} +$

\bigskip
\noindent 
$-6.2337 \cdot Z + 7.87595 \cdot Z \cdot M_{\rm H}$

\bigskip
for $\log T_{max}$ the standard deviation is 3 $\cdot$ 10$^{-3}$. 
Finally, the evolution of the core mass during the whole TP-AGB phase may be estimated
by means of the following formula: 

\bigskip
$M_{\rm H}=M_{\rm H}^*+6.9 \cdot 10^{-3} \cdot k -6.0 \cdot 10^{-5} \cdot k^2$

\bigskip
\noindent
 where
$k=-n,-n+1,....,0,1,....,j$.  $k=0$ corresponds to the last thermal pulse without 
TDU ($M_{\rm H}=M_{\rm H}^*$); 
$n+1$ is the number of TPs of the {\it early phase}, while $j$ is the number of TPs
occurred from the first TDU episode up to the AGB tip ({\it TDU phase + final phase}).
 This formula gives $M_{\rm H}$ within 5 $\cdot 10^{-3}$ $M_\odot$. Then, the 
 envelope mass is easily obtained: $M_{env}=M - M_{\rm H}$, where $M$ is 
 the total mass.

Note that, if the initial mass of the star is too small, the conditions for the activation of the 
third dredge up (namely, $M_{\rm H} \ge 0.60$ $M_\odot$ and $M_{env} \ge 0.4$ $M_\odot$)
are never reached. 
The minimum mass for the occurrence of the TDU, as a function of the mass loss rate, 
are reported in Fig. 2, for solar composition models. 
In stars with masses exceeding this lower limit,
 the envelope composition is modified by the TDU:
$^{12}$C and $s$-elements are enhanced. 
Obviously, a C-star may form, after a suitable number
of TDU episodes, only if the initial mass is large enough. The dashed line in Fig. 2
shows, for solar metallicity models, the lower mass limit for the formation
of a C-star as a function of $\eta$. 

\begin{figure}
\begin{center}
{\centerline{\psfig{file=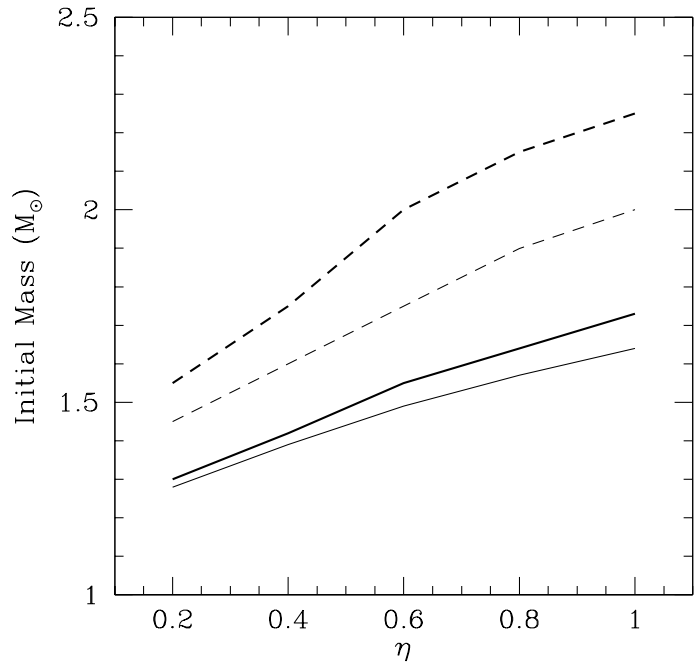,width=12cm}}}
\caption{The minimum initial mass for the occurrence of the TDU (solid lines) 
and the minimum initial mass for C-star progenitors (dashed lines) as a function of the mass loss
rate. The $\eta$ values reported in the abscissa refer to the free parameter in the Reimers formula. 
These results are based on
solar metallicity models ($Z=0.02$ and $Y=0.28$). Thick lines represent the case of models
with minimal dredge up, while thin lines have been derived 
from models that include an exponential decline of the convective velocity at the  
boundaries of the convective zone (see section 3).}
\end{center}
\end{figure}

Both these lower limits depend
on metallicity: the smaller the metallicity, the smaller is the mass for the
activation of the TDU phase and the lower is the minimum initial mass for the 
formation of a C-star.
This occurrence is illustrated in Fig. 3 (for $\eta$=0.5).

\begin{figure}
\begin{center}
{\centerline{\psfig{file=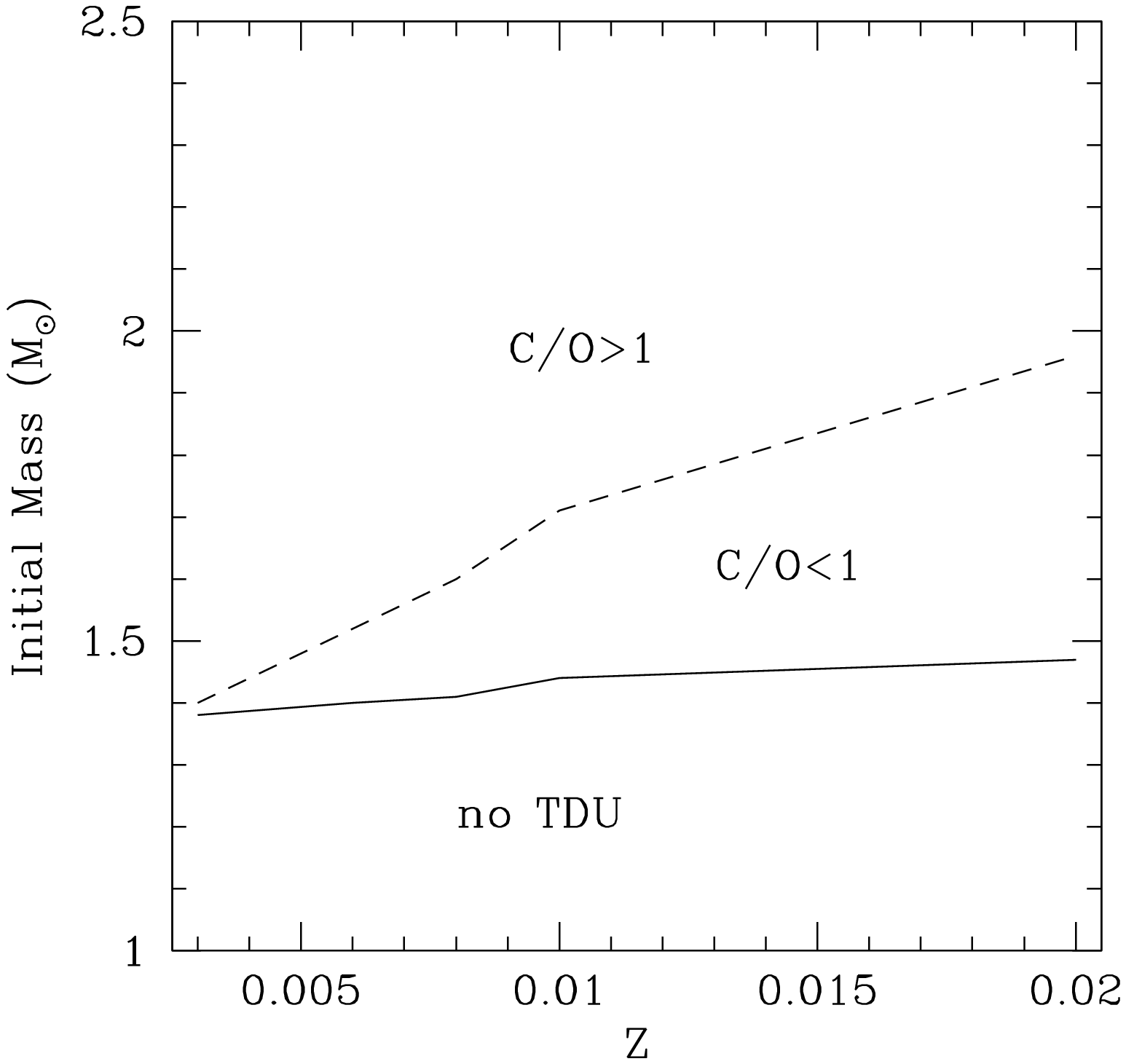,width=12cm}}}
\caption{The minimum initial mass for the occurrence of the TDU (solid line) 
and the minimum initial mass for C-star progenitors (dashed line) as a function of $Z$.
These results are based on models computed with a Reimers mass loss rate ($\eta=0.5$) and
minimal dredge up (see section 3).}
\end{center}
\end{figure}

At low metallicity (i.e. $Z < 4 \cdot 10^{-3}$),
 even taking into account an enhanchement of the $\alpha$-elements (as typically found in
halo stars), the amount of oxygen in the envelope is so low that just one TDU episode
is sufficient to increase the C/O ratio above unity. Correspondingly, the core mass and the 
luminosity at C/O $=$ 1 decrease at low $Z$ (see Fig. 4).

\section{The efficiency of the third dredge up: the Opacity Induced Convection.}

An He-rich zone, as the one left by the shell H-burning,
has a significantly lower opacity than an H-rich zone, 
at the same temperature and pressure. Thus, when the innermost layer of the convective
envelope (H-rich) penetrates into the region of decreasing H
and increasing He (at the epoch of the TDU), the local value of the 
radiative opacity increases.  Since the radiative gradient ($\nabla_{rad}$)
is proportional to the
opacity, the condition $\nabla_{rad} \gg \nabla_{ad}$ (where $\nabla_{ad}$
is the adiabatic temperature gradient) 
takes place at the base of the convective envelope. In this case,
the internal boundary of the convective zone becomes unstable (Becker \& Iben 1979;
Castellani, Chieffi \& Straniero 1990; Frost \& Lattanzio 1996; 
Castellani, Marconi \& Straniero 1998). In fact, even a small perturbation may increase 
the radiative gradient in the formally stable region, 
located immediately below the convective envelope, 
which immediately becomes convectively unstable. 
At present, a satisfactory treatment of this phenomenon has not yet been found.
However, its overall effect on AGB stellar models may be derived
by appling a small perturbation to the boundaries of the convective regions. 
As a tentative scheme, we have supposed that the velocity of the convective elements,
usually evaluated by means of the mixing length theory, does not abruptly drop to 0 at the 
convective boundaries, but
decreases following an exponential decline law (for more details on the mixing algorithm
see section 2 in Chieffi et al. 2001):

\begin{center}
$v=v_{bce}\exp{\left(-\frac{z}{{\beta}H_P}\right)}$,
\end{center}

\noindent
where {\it z} is the distance from the convective boundary, $v_{bce}$ the average 
velocity at the convective boundary, {\it H$_P$} the pressure scale height at the 
convective boundary and $\beta$ is a free parameter that controls 
the steepness of the velocity decline. Usually, when the convective boundary 
is located within a 
chemically homogeneous region,  $v_{bce} \sim 0$ and the convective border is stable.
However,
at the epoch of the TDU, at the base of the convective envelope 
the difference between 
radiative and adiabatic gradients increases. Thus,
$v_{bce}$ increases, causing an inward propagation of the convective instability.
The final result is
a substantial increase of the TDU efficiency. With $\beta=0.1$, for example, 
the amount of H-depleted mass dredged up by convection is about a factor 2 larger than that
found in models without the exponential decline of the convective velocity
(Cristallo et al. 2003).  

Few additional models have been computed by taken $\beta=0.1$ (see last column in Tab. 1). 
The effect on the minimum mass for C-star progenitors is shown in Fig. 2, in the case of 
solar composition models. The corresponding decrease of the minimum C-star luminosity is
illustrated in Fig. 4. 

 \begin{figure}
\begin{center}
{\centerline{\psfig{file=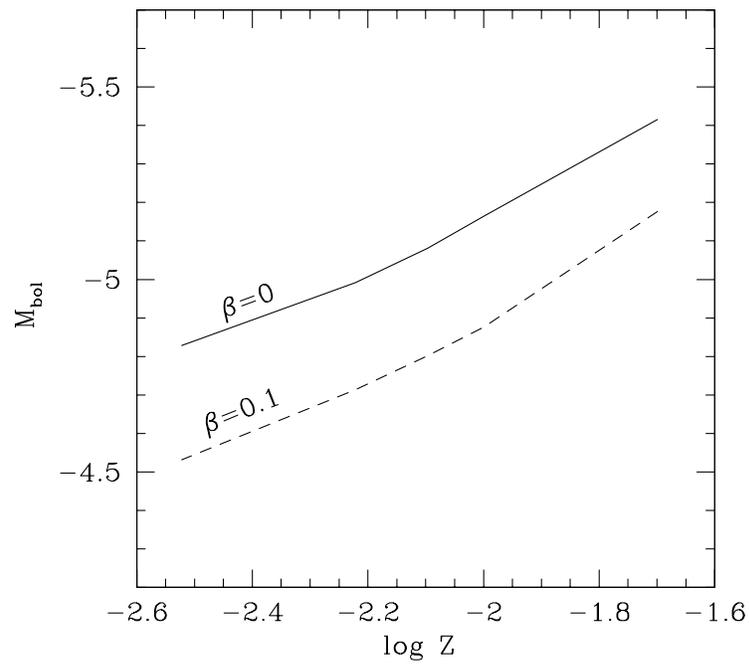,width=12cm}}}
\caption{The minimum luminosity for a C-star as a function of $Z$.
These results are based on models computed with a Reimers mass loss rate ($\eta=0.5$) and
with (dashed line) and without (solid line) exponential decline of 
the convective velocity.}
\end{center}
\end{figure}



\begin{table}
\begin{center}
\begin{tabular}{|c|c|c|c|c|}
  \hline
$M$ ($M_\odot$) & $Y$ & $Z$ & $\eta$ & $\beta$ \\
 \hline
1.0 & 0.280 & 0.020  & 0   & 0    \\
1.5 & 0.280 & 0.020  & 0   & 0\\
1.5 & 0.280 & 0.020  & 0.4 & 0/0.1 \\
2.0 & 0.280 & 0.020  & 0.4 & 0/0.1 \\
3.0 & 0.280 & 0.020  & 0   & 0 \\
3.0 & 0.280 & 0.020  & 1.5 & 0/0.1 \\
1.5 & 0.255 & 0.006  & 0.5 & 0/0.1 \\
3.0 & 0.255 & 0.006  & 1.5 & 0     \\
1.5 & 0.230 & 0.003  & 0.4 & 0/0.1 \\
3.0 & 0.230 & 0.003  & 1.5 & 0\\
\hline
\end{tabular}
\caption{The computed grid of models. $M$ is the initial mass. $Y$ and $Z$ are the
 initial helium content and the metallicity, respectively. $\eta$ is the free parameter
 in the Reimers mass loss formula. $\beta$ is the free parameter in the exponential decline law
 for 
  the boundary velocity of the convective elements (section 3).}
\end{center}
\end{table}

\section{Final remarks}

In this paper, based on an homogeneous grid of low mass 
AGB stellar models, we provide some interpolation formulae  
for basic ingredients used in nucleosynthesis calculations.
These formulae are strictly valid for $M \le 3 M_\odot$, at solar metallicity, and 
for $M \le 2.5 M_\odot$,  in the metallicity range $0.003 \le Z < 0.02$. 

 A study of the third dredge-up as a function of 
 total mass and metallicity  has been recently published by 
Karakas, Lattanzio \& Pols (2002), while a comparison among the AGB stellar models computed by means of  
 different evolutionary codes has been discussed by Lugaro et al. (2003). Once the same mass loss
 rate is adopted, a substantial  
  agreement is found between our (FRANEC) results  and those 
   obtained by using the Mount Stromlo Stellar Structure Program (Frost and Lattanzio, 
   1996; Karakas, Lattanzio \& Pols 2002).
   
As a final remark, note that although we have used a Reimers formula to evaluate the mass loss, since the 
 fittings formulae are given as a function of the envelope and core masses,
  they can be used for any mass loss law.

%

\section*{Acknowledgments}

This research was supported by the Italian grant MURST-FIRB,
by the Spanish grant AYA2002-04094-C3-03 and
by the Andalusian grant FQM-292.

\section*{References}

\reference Abia, C., Busso, M., Gallino, R., Dominguez, I., Straniero, O., \& Isern, J. 2001, ApJ 559, 1117
\reference Abia, C., Dominguez, I., Busso, M., Gallino, R., Masera, S., Straniero, O.,
de Laverny, P., Plez, B., \& Isern, J. 2002, ApJ 579, 817
\reference Becker, S.A., \& Iben, I. Jr. 1979, ApJ 232, 831
\reference Busso, M., Gallino, R., Lambert, D.L., Travaglio, C., \& Smith, V.V. 2001, ApJ 557, 802
\reference Busso, M., Gallino, R., \& Wasserburg, G.J. 1999, ARAA 37, 239
\reference Castellani, V., Chieffi, A., \& Straniero, O. 1990, ApJS 74, 463
\reference Castellani, V., Marconi, M., \& Straniero, O., 1998, AA 340, 160 
\reference Chieffi, A., Dominguez, I., Limongi, M., \& Straniero, O. 2001, ApJ 554, 1159
\reference Cristallo, S., Gallino, R., \& Straniero, O., 2003, Mem. SAIt (in press).
\reference Frost, C., \& Lattanzio, J.C. 1996, ApJ 473, 383
\reference Fusi Pecci, F., \& Renzini, A. 1976, AA 46, 447
\reference Gallino, R., Arlandini, C., Busso, M., Lugaro, M., Travagio, C., Straniero, O.,
Chieffi, A., \& Limongi, M. 1998, ApJ 497, 388
\reference Gallino, R., Busso, M., \& Lugaro, M. 1997, in {\it
Astrophysical Implications of the Laboratory Study of Presolar Material},
ed. T. J. Bernatowicz and E. Zinner, {\it AIP Conf. Proc.}, 402,
(Woodbury: American Institute of Physics), 115
\reference Karakas, A.I., Lattanzio, \& Pols, O.R. 2002, PASA 19, 515
\reference Lambert, D.L., Smith, V.V., Busso, M., Gallino, R., \& Straniero, O. 1995, ApJ 450, 302
\reference Lugaro, M., Herwig, F., Lattanzio, J.C., Gallino, R., \& Straniero, O. 2003, ApJ 586, 1305
\reference Reimers, D. 1975, in {\it Problems in Stellar Atmospheres and Envelopes}, 
ed. B. Baschek, W.H. Kegel and G. Traving, (Berlin Springer), 229
\reference Straniero, O., Gallino, R., Busso, M., Chieffi, A., Limongi, M., \& Salaris, M.
1995, ApJ 440, L85
\reference Straniero, O., Chieffi, A., Limongi, M., Gallino, R., Busso, M., \& Arlandini, C.
1997, ApJ 478, 332

\end{document}